\documentclass[journal]{IEEEtran}
\usepackage{cite}
\usepackage{amssymb}
\usepackage{amsmath}
\usepackage{amsfonts}
\usepackage{color}
\usepackage{algorithmic}
\usepackage{algorithm}
\usepackage{amsthm}
\usepackage{graphicx}
\usepackage{epstopdf}
\usepackage{booktabs}
\usepackage{lipsum}
\usepackage{cuted}
\usepackage{textcomp}
\usepackage{verbatim}
\usepackage{subcaption}
\usepackage{array}  
\usepackage{float}

\usepackage{hyperref}

\hypersetup{hidelinks,
	colorlinks=true,
	allcolors=black,
	pdfstartview=Fit,
	breaklinks=true}
\ifCLASSINFOpdf

\else

\fi

\begin{document}
	\setlength{\abovedisplayskip}{2pt}
	\setlength{\belowdisplayskip}{2pt}
	\title{Split Federated Learning Empowered Vehicular Edge Intelligence: Concept, Adaptive Design and Future Directions}
 	
	\author{Xianke Qiang, Zheng Chang,~\IEEEmembership{Senior~Member,~IEEE,} Chaoxiong Ye, Timo H\"am\"al\"ainen,~\IEEEmembership{Senior~Member,~IEEE}, Geyong~Min,~\IEEEmembership{Senior Member,~IEEE}
		
		\thanks{X. Qiang and Z. Chang are with School of Computer Science and Engineering, University of Electronic Science and Technology of China, Chengdu 611731, China. Z. Chang, C. Ye and T. H\"am\"al\"ainen are with Faculty of Information Technology, University of Jyv\"askyl\"a, P. O. Box 35, FIN-40014 Jyv\"askyl\"a, Finland. G. Min is with Department of Computer Science, University of Exeter, Exeter, EX4 4QF, U.K.}

 }
	\maketitle

\begin{abstract}
    To achieve ubiquitous intelligence in future vehicular networks, artificial intelligence (AI) is essential for extracting valuable insights from vehicular data to enhance AI-driven services. By integrating AI technologies into Vehicular Edge Computing (VEC) platforms, which provides essential storage, computing, and network resources, Vehicular Edge Intelligence (VEI) can be fully realized. Traditional centralized learning, as one of the enabling technologies for VEI, places significant strain on network bandwidth while also increasing latency and privacy concerns. Nowadays, distributed machine learning methods, such as Federated Learning (FL), Split Learning (SL), and Split Federated Learning (SFL), are widely applied in vehicular networks to support VEI. However, these methods still face significant challenges due to the mobility and constrained resources inherent in vehicular networks. In this article, we first provide an overview of the system architecture, performance metrics, and challenges associated with VEI design. Then, the adaptive design of SFL, namely Adaptive Split Federated Learning (ASFL) is introduced. The proposed ASFL scheme dynamically adapts the cut layer selection process and operates in parallel, optimizing both communication and computation efficiency while improving model performance under non-IID data distribution. Finally, we highlight future research directions to shed the light on the efficient design of SFL.
\end{abstract}

\begin{figure*}[t]
    \centering
    \includegraphics[width=1.0\textwidth]{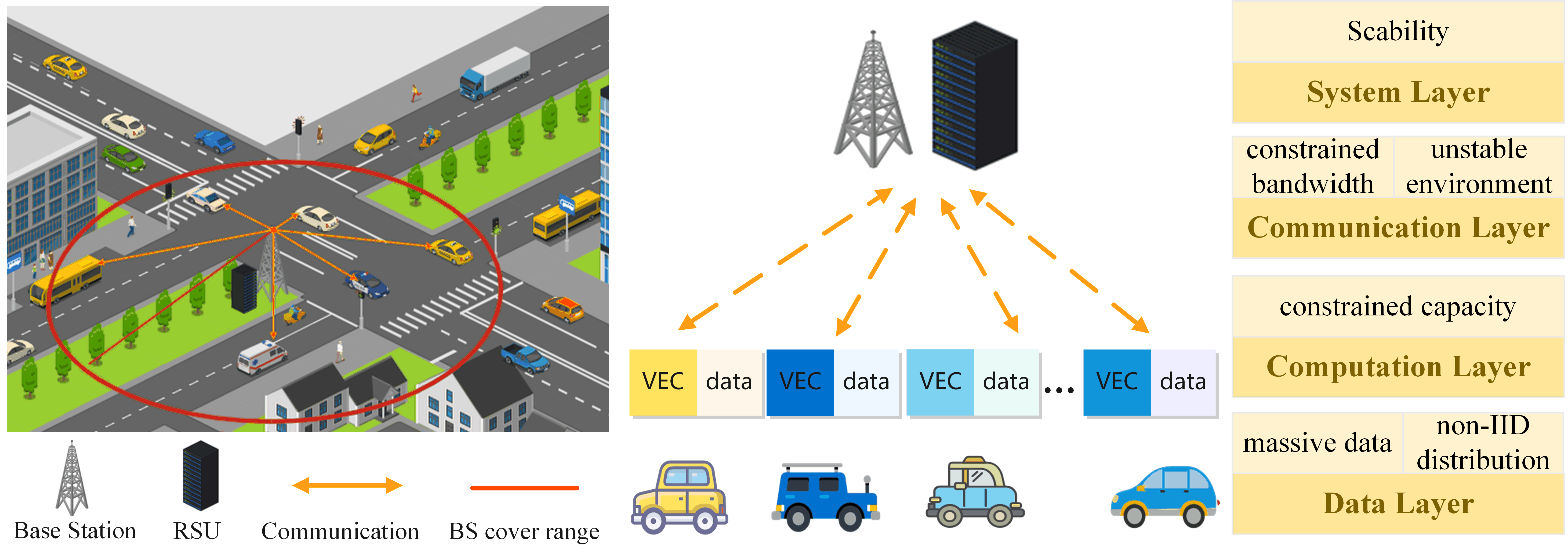} 
    \caption{VEI system architecture.}
    \label{fig:architecture} 
\end{figure*} 
\section{Introduction}
    { The Intelligent Transportation System (ITS) has emerged as a promising approach to enhancing transportation safety, efficiency, and autonomy, driven by advances in wireless communications and the Internet of Things (IoT). By integrating AI technologies into Vehicular Edge Computing (VEC) platforms, which provides essential storage, computing, and network resources, Vehicular Edge Intelligence (VEI) can be fully realized. Recently, researchers are focusing on VEI as a crucial element for advancing ITS\cite{10133894}. Machine Learning (ML) methods have shown significant promise in various ITS applications, such as object detection, traffic sign classification, congestion prediction, and velocity/acceleration forecasting\cite{ye2018machine}. However, the traditional approach of transmitting raw data to centralized servers for ML processing introduces considerable privacy risks and demands substantial bandwidth for wireless communication.}\par
    With increasing emphasis on privacy and widespread deployment of edge computing in vehicular networks, Federated Learning (FL) emerges as a promising distributed learning framework for implementing VEI. FL enables the vehicles to train the local model with private data, and then upload the local model for Road Side Unit (RSU) to aggregate. Despite the potential of FL in VEI, there remain numerous challenges \cite{zhang2023federated}. One significant problem is the high heterogeneity among the clients involved in training \cite{yang2022flash}. { Differences in local data cause the updates from each client to become nearly orthogonal, which weakens the global updates and slows down the model’s convergence\cite{an2023federated}. }Another primary concern of FL is how to protect user privacy since sensitive information can still be revealed from model parameters or gradients by a third-party entity or the RSU\cite{shen2023ringsfl}. Furthermore, with the development of AI, we have entered the era of large models, which are progressively growing in size and complexity. Training complete and large models on resource-constrained vehicles poses a significant challenge. \par
    Meanwhile, Split Learning (SL) is also one of the underlying technologies for achieving VEI, where the whole AI model (e.g., CNN) is partitioned into several sub-models (e.g., a few layers of the entire CNN) with the cut layer and distributing them to different entities (e.g., the vehicle-side model at the vehicles or the RSU-side model at the RSU)\cite{liu2022wireless}. By offloading computation-intensive portions to the RSU and preserving privacy-sensitive portions locally, SL can significantly reduce the computation load of model training on resource-constrained devices, and has great potential to empower the future ITS. However, utilizing the sequential SL directly may induce extra communication load and time delay.\par 

    It is worthy noticing that a novel framework called Split Federated Learning (SFL) combines the ideas of SL and FL to parallel the training process \cite{thapa2022splitfed}. As for SFL in vehicular network, the vehicle downloads the vehicle-side model and executes forward propagation to upload the smashed data to the RSU. Then the RSU performs the forward and backward propagation with received smashed data, and broadcasts the gradients of smashed data. After that, the updated vehicle-side model is upload to RSU for aggregation. SFL not only reduces communication load and latency comparing with SL, but also reduces vehicle computation load, which makes it more suitable for VEC systems. Firstly, SFL enables the vehicles to participate in training without compromising data privacy, thus reducing the risk of data leakage. It enhances vehicle engagement and provides more data. Secondly, by offloading part models to RSU, SFL alleviates the computational bottleneck of vehicles. Thirdly, such a parallel design greatly enhances the scalability of SFL schemes compared to SL, allowing the system to accommodate more vehicles within the communication range of RSU, especially in high-speed mobility scenarios. These advantages shed the light on emerging applications such as cooperative autonomous driving and intelligent traffic navigation. \par 

    \begin{figure*}[htbp]
        \centering
        \includegraphics[scale=0.7]{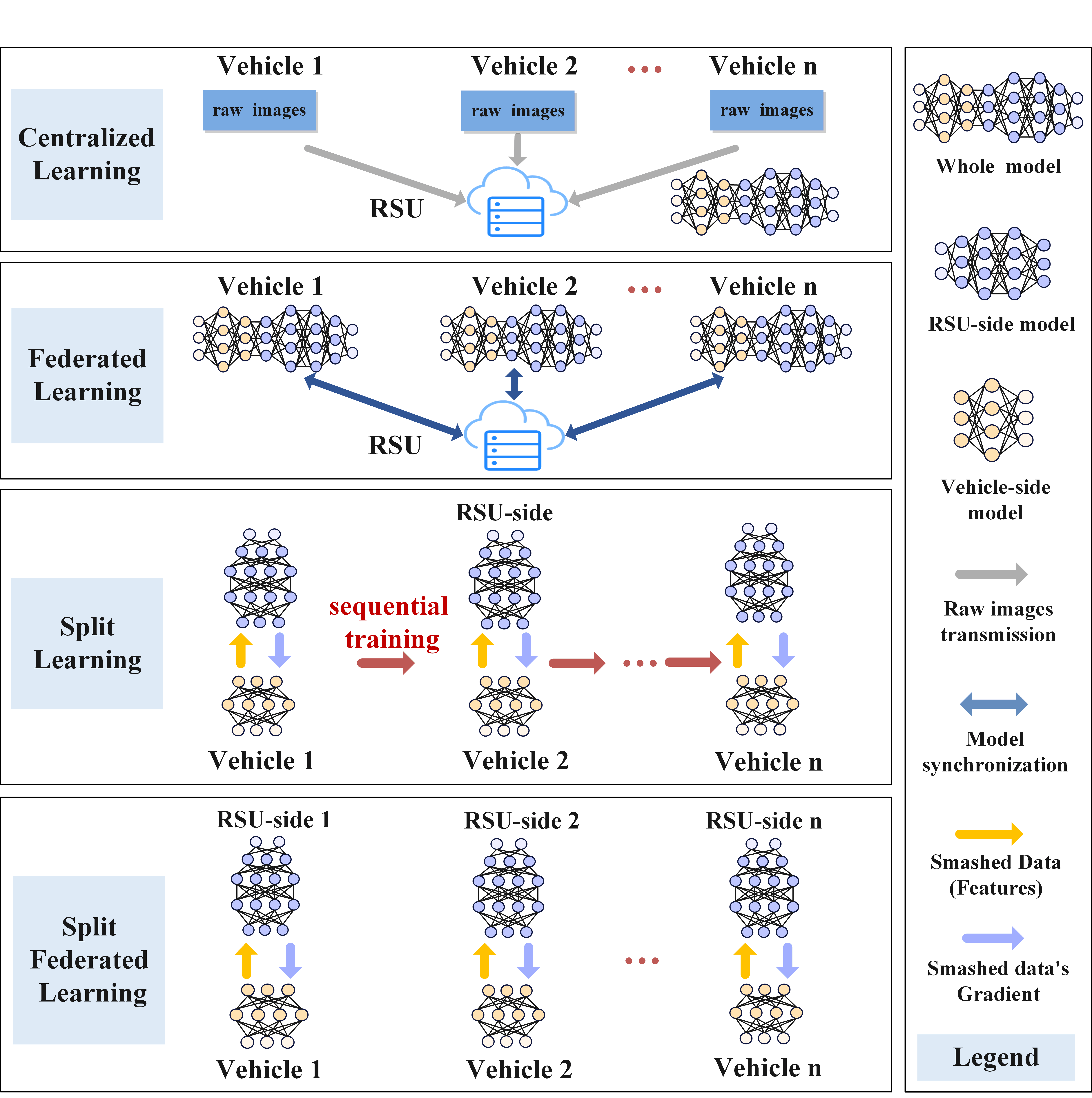}
        \caption{The workflow of centralized learning, federated learning, split learning and split federated learning.}
        \label{fig:workflow}
    \end{figure*}

    However, SFL still faces new challenges when it is applied to VEI due to the mobility and constrained resources of vehicular network. Firstly, continuously moving vehicles may drive out of the RSU's communication range during the training process, leading to interruptions in training. Thus, how to select as many vehicles as possible that can successfully transmit data to participate in training becomes a critical issue. Secondly, there is a significant difference in the computing capabilities of different vehicles. Choosing different cut layers affects system latency, energy consumption, and even privacy. For different vehicle-side models, selecting optimal cut layer to minimize latency, energy consumption, and maximize privacy becomes another important challenge. Thirdly, compared to FL, SFL offloads part of model to RSU, reducing computation load on the vehicle side but increasing communication load, essentially trading communication time for computational time. Since a large number of vehicles participate in training in a single round, balancing system computation latency and communication latency to minimize overall system latency is also an important consideration. \par
    { Motivated by the necessities of cut layer selection, in this article, we introduce Adaptive Split Federated Learning (ASFL) scheme as an advanced solution to address the limitations of traditional SFL in VEC systems. ASFL adapts dynamically to unstable network conditions and vehicle capabilities, optimizing the selection of cut layers and balancing the trade-offs between computation and communication. This work represents an early attempt to provide a comprehensive overview of SFL-empowered VEI.}
    
    The rest of this paper is structured as follows. We begin by presenting the background information, including system architecture, performance metrics, and challenges specific to VEI, with a particular focus on distributed implementations. Next, we introduce the novel parallel and adaptive ASFL scheme as an enabling technology for VEI and provide a case study to evaluate its performance in a real communication environment. Finally, we discuss open research directions for future work.

\section{Vehicular Edge Intelligence: Architecture, Performance Metrics and Challenges}

In this section, we will firstly introduce system architecture shown in Fig. \ref{fig:architecture}. And then we introduce intelligent metrics of vehicular network systems from the aspects of training and testing, time and energy, privacy and security. Then we analyze the facing challenges in the distributed implementation of VEI.

\subsection{System Architecture}

VEI utilizes the computation and communication resources of vehicles, combined with AI technologies. VEI relies on the effective utilization of extensive data gathered from numerous vehicles for model training. We can categorize the implementation of VEI into four approaches: centralized machine learning (CL), and distributed collaborative methods including FL, SL, and SFL. The details of these four approaches are shown in Fig. \ref{fig:workflow}.\par

CL aggregates training data at centralized locations, such as cloud data centers. However, transmitting vast amounts of vehicular data to these centers strains network bandwidth and further exacerbates latency issues. Additionally, vehicular data often encompasses sensitive information, including personal data related to user information (e.g., license plate numbers, facial features, and vehicle details). Thus, the imperative to retain data on local devices emerges to safeguard user privacy.\par 

Distributed collaborative learning emerges as a promising technical method under exploration by the research community to tackle the challenges of CL. Generally, distributed collaborative learning entails the joint training of a global model through collaboration, without directly access to the decentralized raw data. This approach holds significant appeal for applications seeking to leverage the wealth of data widely distributed in vehicles. Notably, FL and SL stand out as two representative and emerging methods within the realm of distributed collaborative learning.\par

FL allows many data owners to work together to train a shared AI model without revealing their individual data. In FL, each vehicle independently trains a local AI model using its own data. {Then these local models are collected by the RSU to aggregate a global model, preserving data privacy by keeping the data on the local vehicles. It's a promising solution for dealing with data privacy challenges in Internet of Vehicles (IoV). However, FL requires each vehicle to have sufficient resources for training AI models, which can be challenging for resource-limited vehicles, especially when dealing with complex models like deep neural networks.}\par

{SL is another distributed collaborative learning approach, the whole model is partitioned to be collaboratively trained at vehicles and the RSU. The SL operates in three main steps. Initially, the vehicle downloads the vehicle-side model and performs forward propagation to transmit the smashed data (the forward propagation output of cut layer) to RSU. Subsequently, the RSU conducts forward and backward propagation of RSU-side model with the smashed data as input and the gradient of smashed data as output, then sends the gradient of smashed data back to vehicle for vehicle-side model backward propagation and updating. Next, the updated vehicle-side model is transferred to the next vehicle to repeat the above process until all vehicles are trained. SL allows vehicles to offload part of whole model to RSU thus making it possible to leverage flexible resource management in computing for supporting model training. However, the sequential vehicle-RSU collaboration in SL limits its capability of involving vast data distributed in massive vehicles for model training.\par

Combining the advantages of SL and FL, SFL can not only allows vehicles to offload part of training model to RSU, but also enables parallel training. SFL mainly constraints three steps. Firstly, all vehicles downloads the vehicle-side model from RSU and execute forward propagation to upload the smashed data to the RSU. Afterward, the RSU-side model perform the forward and backward propagation with received smashed data, and then send the gradients of smashed data to the vehicles for update respectively. Finally, the updated vehicle-side models are upload to RSU for aggregation.}\par

\subsection{VEI Performance Metrics}
\subsubsection{Training $\&$ Testing}
{For AI models, performance is the fundamental criterion for evaluating their effectiveness. Training accuracy is a critical indicator of the model's learning capacity, reflecting its performance on the training data. Testing accuracy, in contrast, is a key metric for evaluating the model's generalization capability, reflecting its performance on previously unseen data. Furthermore, the convergence of the AI model is an essential consideration, as it reflects the stability and efficiency of the training process, ensuring that the model reaches an optimal state. Therefore, for AI models within the vehicular network, besides considering challenges such as computational power and wireless resource allocation, it is essential to focus on model performance to ensure their learning, generalization and convergence capabilities especially in the case of non-IID distribution of data. }

 \subsubsection{Time $ \& $ Energy}
    In vehicular networks, tasks often have high time sensitivity due to the high-speed mobility of vehicles.  Given that different vehicles stay within the communication range of RSUs for a short and uncertain period, the overall system's model training time becomes a crucial metric. Additionally, energy consumption serves as a significant evaluation criterion. Vehicles are primarily served as transportation tools. While utilizing onboard data for training can enhance the passenger experience, it is crucial to avoid excessive energy consumption during model training to ensure it does not compromise the vehicle's primary transport function. Therefore, energy consumption should also be taken into consideration.
    
\subsubsection{Privacy $\&$ Security}
    With growing concerns regarding data privacy, especially in connected vehicles, data leakage not only compromises property security but also poses more severe risks to personal safety and traffic safety. Privacy regulations, such as the EU's General Data Protection Regulation (GDPR) or the California Consumer Privacy Act (CCPA), impose limitations on the collection and immediate utilization of users' sensing or perception data for model training and inference \cite{lyu2023scalable}. The local gradients uploaded by clients could potentially be exploited by attackers to infer the data samples of membership\cite{9109557}. In SL, the data owner and the label owner also have privacy issues, although they only share the intermediate data, i.e., the smashed data and the cut layer gradients. The smashed data (i.e., the extracted features) shared from the data owner may be used by the attacker to reconstruct its training data \cite{he2020attacking}. To alleviate the burden of privacy protection on the network side, the model training framework necessitates retaining raw data within local vehicles or using some privacy technology such as differential privacy technology \cite{10229176}.

    \begin{figure*}[htbp]
        \centering
        \includegraphics[scale=0.9]{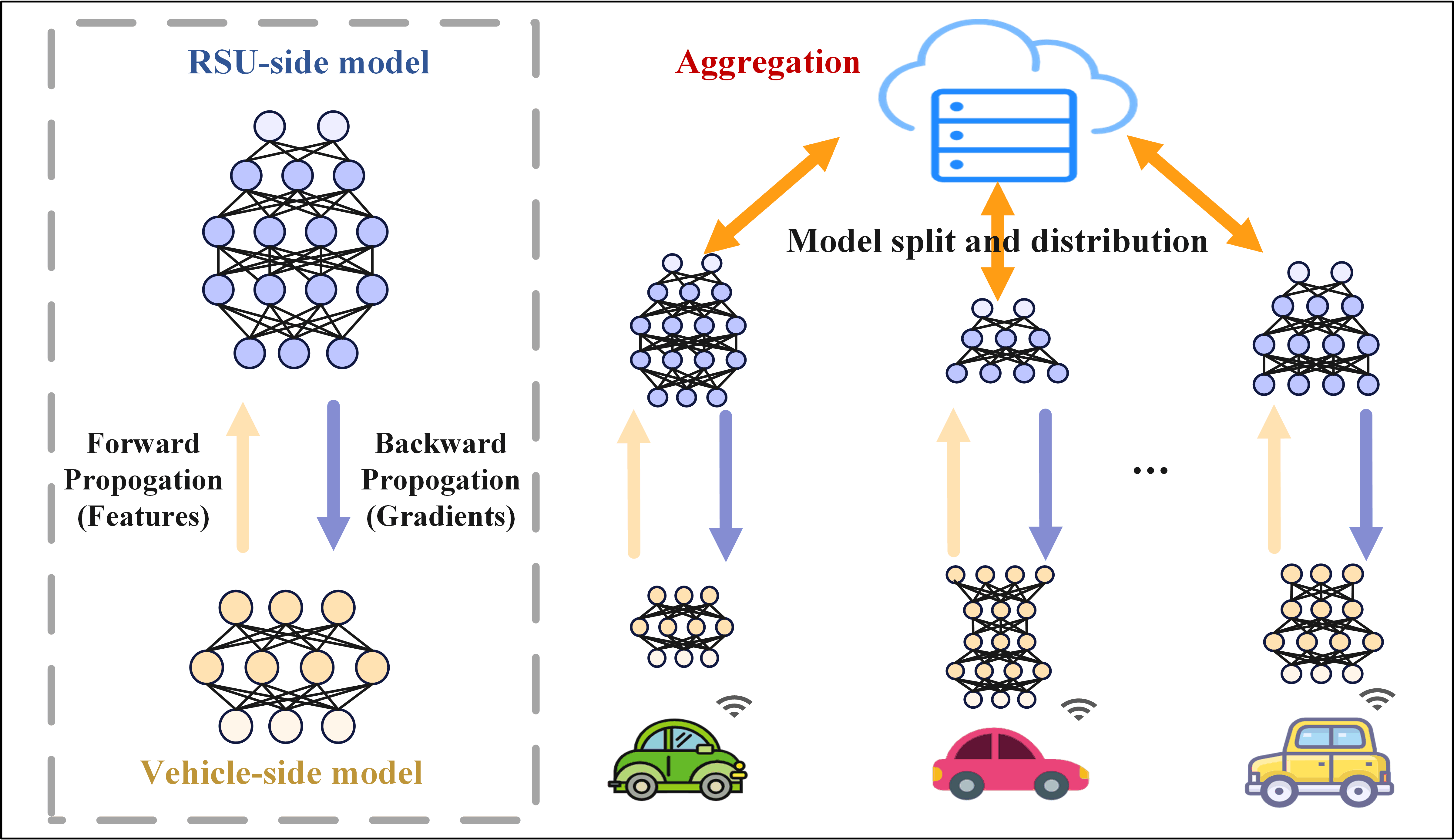}
        \caption{The workflow of adaptive split federated learning.}
        \label{fig: propose} 
    \end{figure*}
    
\subsection{Challenges}
In this subsection, we focus on the challenges faced by distributed machine learning in implementing VEI from different layers.

\subsubsection{Data Layer}
The intelligence of VEI is derived from the significant volume of data engaged in its training process obtained from vehicles. From data layer perspective, the vehicular network primarily encounters two challenges to achieve VEI. Firstly, a large amount of devices have massive amounts of data, which means more energy is required for training, and not all data is useful, as even many data are redundant. Secondly, in the IoV, the data involved in training exhibits high heterogeneity, which means they may originate from different types of vehicles, various geographical locations, and diverse driving conditions. This heterogeneity poses significant challenges to model training because models need to effectively capture and adapt to these variations. 

\subsubsection{Computation Layer}
Computational capability is crucial for training more complex AI models to provide intelligent services, particularly as we enter the era of large model. However, in the context of time-sensitivity, the task of training AI models on vehicles with restricted computing power can prove excessively time-consuming. Moreover, the heterogeneity of computing platforms presents a significant challenge, stemming from the diverse types and computational capabilities of vehicles, thereby inducing instability in system operations. In addition, the large AI models does indeed bring challenges. Traditional FL requires training the entire model on each device, which may not be realistic for vehicles with limited resources. In fact, running large models on vehicles is not economical because it consumes a lot of energy. The primary task of vehicles is intelligent transportation, so sacrificing too much energy needed for transportation in favor of model training is unnecessary. 

\subsubsection{Communication Layer}
In distributed learning, high-performance wireless networks play a crucial role in accelerating the implementation of VEI, as intermediate parameters of model training need to be transmitted through wireless network. For instance, in FL, models are uploaded in parallel to the RSU for aggregation, while in SL, smashed data and their corresponding gradients must be communicated between vehicles and the RSU via wireless networks. However, owing to the inherent instability of wireless networks, the extensive participation of vehicles in training, and the varying distances between vehicles and the RSU, not all vehicles can access sufficient communication bandwidth. Moreover, due to the high-speed movement of vehicles, some may exit the communication range of the RSU during training, hindering model completion. 

\subsubsection{System Layer}
Different designs in training architectures have multifaceted impacts on the implementation of VEI. Choosing an appropriate training architecture not only affects the system's performance and efficiency but also involves aspects such as privacy, security, scalability, and flexibility. A centralized training architecture may lead to increased burdens in data transmission and computation, resulting in system latency and energy consumption increasing, as well as posing risks of data leakage. Conversely, a distributed training architecture offers better privacy protection and security, as well as improved scalability and flexibility. Therefore, selecting the right training architecture is crucial, requiring comprehensive consideration of multiple factors to ensure the efficient and secure operation of the VEI system. In traditional SL, the RSU with RSU-side models must sequentially serve vehicles with unique local datasets. Due to this sequential training process, the overall latency for each training epoch increases linearly with the number of vehicles. This increased latency could hinder the scalability of split learning, particularly in large-scale ITS vehicle deployments.

\section{Adaptive and Parallel Split Federated Learning}
Bearing in mind the aforementioned challenges, such as non-IID distribution of data, shortage of computing resources, tight communication resources, and low system scalability, this section introduces an Adaptive Split Federated Learning (ASFL) scheme. The system can dynamically select the cut layer based on environmental conditions for every vehicle. {Compared to SL, the proposed scheme reduces communication load, and compared to FL, it reduces computation load. This makes the distributed learning more adaptable to the mobility characteristics of the vehicular network.}

\subsection{Adaptive Split Federated Learning Scheme}
We consider a general vehicular network that includes one RSU and a set of $N$ vehicles. The data set of the vehicle $n$ is denoted as $\mathcal{D}_n$, and $\left|\mathcal{D}_n\right|$ is the number of training data samples of vehicle $n$. \par

The objective is to collaboratively train a global AI model that minimizes the global loss function based on the global dataset collected from all vehicles:
        \begin{equation}
            \min_{\boldsymbol{\omega}} L(\boldsymbol{\omega})=\frac{1}{\sum_{n=1}^{N}  \left|\mathcal{D}_n\right|}\sum_{n=1}^N { \left|\mathcal{D}_n\right|} L_n(\boldsymbol{\omega}),\nonumber
        \end{equation} 
    where $L_n(\boldsymbol{\omega})$ denotes the local loss function of vehicle $n$.\par
    The full model of vehicle $n$ in the $t$-th round $\boldsymbol{\omega}_{t}^{n,\epsilon}$ includes two non-overlapping sub-models with $\epsilon$-th cut layer, represent as vehicle-side model $\boldsymbol{\omega}^{V,\epsilon}_{t}$ and RSU-side model $\boldsymbol{\omega}^{S,\epsilon}_{t}$, it can be denoted by  $\boldsymbol{\omega}_{t}^{n,\epsilon} = \{\boldsymbol{\omega}^{V,\epsilon}_{t};\boldsymbol{\omega}^{S,\epsilon}_{t}\},$
        and the global model update principle is as follows:
            \begin{align}
               \boldsymbol{\omega}_{t+1} = \boldsymbol{\omega}_t -\sum_{n=1}^{N}   \frac{1}{N}(\boldsymbol{\omega}_{t+1}^{n,\epsilon} - \boldsymbol{\omega}_{t}),\nonumber
            \end{align}
\subsection{Workflow of ASFL} 
The workflow of ASFL is shown in Fig. \ref{fig: propose}. Firstly, the RSU chooses different cut layers and distributes different vehicle-side model to different vehicles according to cut layer selection strategy. Secondly, the vehicles train vehicle-side model they received over the local datasets parallel, and then send the corresponding smashed data, to the RSU. Thirdly, the RSU is supposed to have sufficient resources and can provide powerful computing capability, such that it sequentially performs forward propagation to the RSU-side model with the received smashed data to calculate the loss function respectively, and then broadcasts the gradients of smashed data. Finally, vehicles update their vehicle-side models and upload them to the RSU, where they are merged with the corresponding RSU-side models to form a whole model. These whole models are then aggregated to achieve a global model update. \par

\begin{figure*}[htbp]
    \centering
    \includegraphics[width=0.8\textwidth]{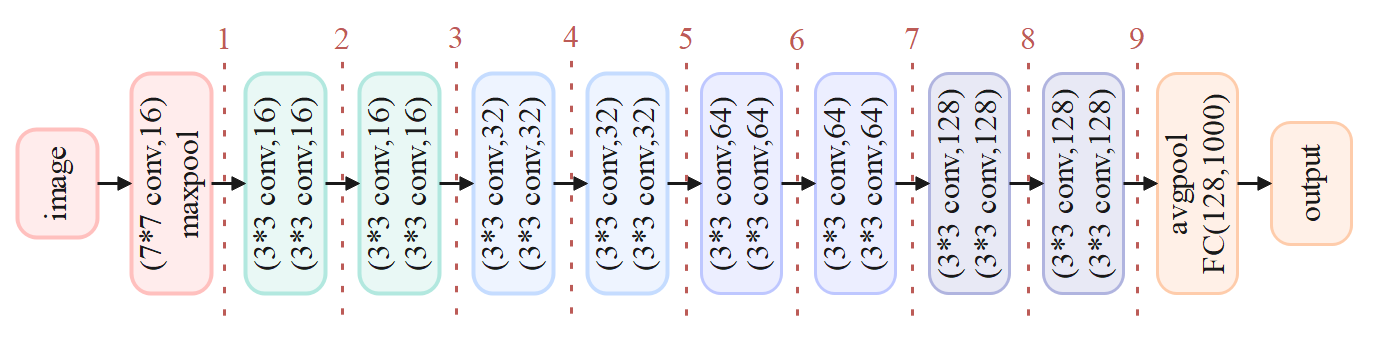} 
    \caption{ResNet18 Model Structure.}
    \label{fig: cut} 
\end{figure*}

{\subsection{Cut Layer Selection Strategy}
    In high-mobility vehicular scenarios, vehicles are constantly moving, and the wireless transmission channel environment is unstable. Additionally, differences in sensors and hardware across vehicles lead to significant variations in data and computation capacities. To address these challenges and improve the feasibility of adaptive cut layer selection algorithms, we propose a relatively simple cut layer selection strategy based on transmission rate. This strategy primarily aims to explore the adaptability and necessity of such algorithms, with a focus on minimizing communication latency and providing a foundation for future research.\par
    In our case study, we assume that all vehicles remain within the RSU communication range for the same duration. Therefore, we consider only the transmission rate of different vehicles when selecting the cut layers. As shown in Fig. \ref{performance}(\subref{communication_overload}), the communication load of SL is substantial, as it requires not only the transmission of vehicle-side model but also the exchange of intermediate messages(smashed data and the grandients) for model training. Compared to FL, SL and SFL methods can reduce computation load by offloading part of the training model to the server. Essentially, SFL increases the communication load to alleviate the computation load compared to FL. ASFL can dynamically adjust the balance between communication and computation load. As the cut layer increases (e.g., SFL2, SFL4, SFL6, SFL8), the increase in communication load relative to FL gradually decreases, as does the reduction in computation load. Therefore, when the transmission rate between the vehicle and the RSU is high, we consider selecting a smaller cut layer, thereby increasing the communication load more to reduce the computation load more. Conversely, when the transmission speed is slower, indicating a congested network, we opt for a later cut layer, thereby adding less communication load to achieve a smaller reduction in computation load.\par

    We use ResNet18 as the training AI model, which has a total of 9 split points, as illustrated in Fig. \ref{fig: cut}. The cut layer selection strategy is defined as follows:

        \begin{equation}
            \label{cutselection}
            c_{n} = 
            \begin{cases}
            8, &        0  < r_n^t \leq \bar{R_1}\\
            6, & \bar{R_1} < r_n^t \leq \bar{R_2}\\
            4, & \bar{R_2} < r_n^t \leq \bar{R_3}\\
            2, & \bar{R_3} < r_n^t \leq \bar{R_4}\\
            \end{cases}
            \nonumber
        \end{equation}
    where $r_n^t$ is the transmission rate of vehicle $n$ in the $t$-th round, and $\bar{R_1} \leq \bar{R_2} \leq \bar{R_3} \leq \bar{R_4}$ are the threshold speeds.}

\subsection{Performance Evaluation}
 \subsubsection{Experiment Setting}
    {In our experiment, we utilized the NVIDIA GeForce RTX 3060 GPU as the RSU, while the NVIDIA GeForce RTX 3060 CPU served as vehicles. In total, there are four vehicles and one RSU, and we use sockets to communication between vehicles and servers.} \par
    The learning rate is 0.0001, and the batch size is 16, local epochs is 5. We consider three baselines, FL, SL and SFL. The number of SFL2,4,6,8 means the number of cut layer. We using CIFAR10 \cite{krizhevsky2009learning} as our simulation dataset. Notably, data distribution at vehicles is non-IID, which widely exists in practical systems. To capture the heterogeneity among mobile vehicles in these datasets, we impose a constraint where each vehicle retains only six out of the ten possible labels, with sample sizes varying according to a power law as described in \cite{li2020federated}. The official implementations of ASFL are available at \cite{github_code}.

\begin{figure*}[htbp]
    \begin{minipage}{0.48\linewidth}
        \centering
        \begin{subfigure}{0.9\linewidth}
             \includegraphics[width=\textwidth]{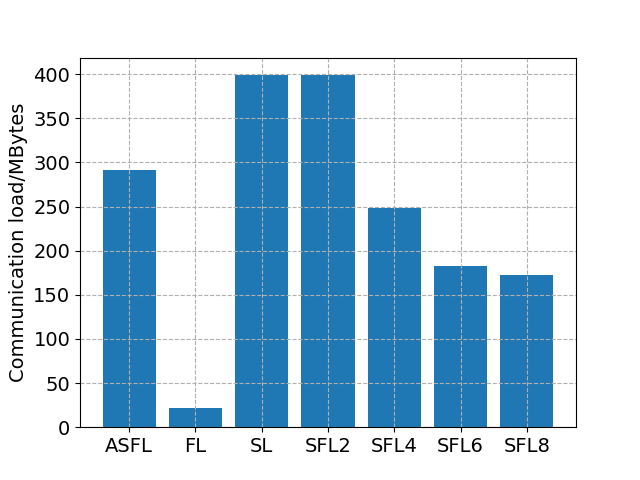}
            \caption{Communication load}
            \label{communication_overload}
        \end{subfigure}
         \begin{subfigure}{0.9\linewidth}  
            \includegraphics[width=\textwidth]{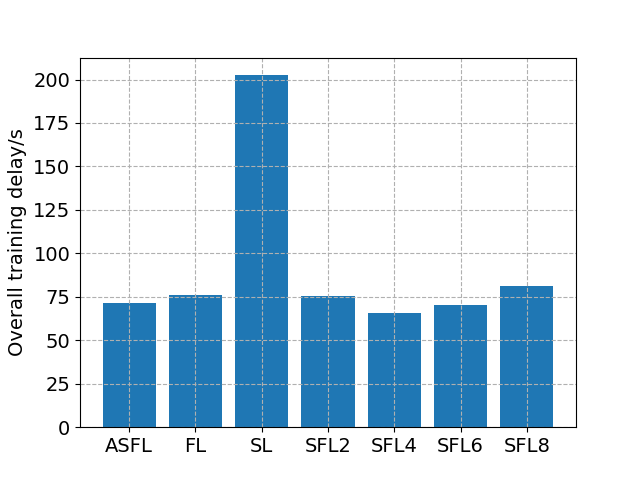}
            \caption{Overall training delay}
            \label{communication_time}
        \end{subfigure}
    \end{minipage}
    \begin{minipage}{0.48\linewidth}
        \centering
        \begin{subfigure}{0.9\linewidth}
            
             \includegraphics[width=\textwidth]{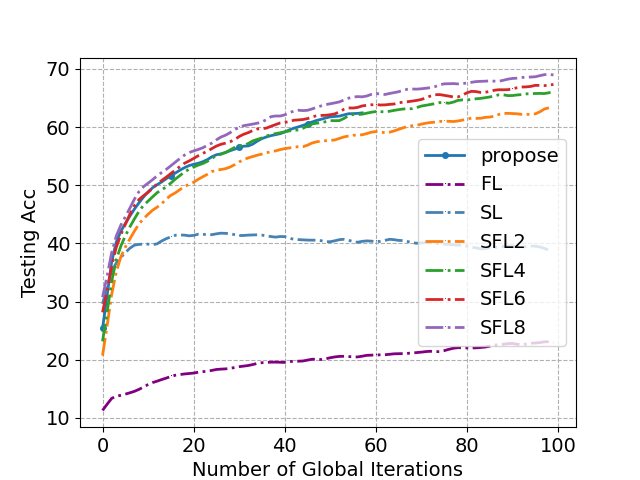}
            \caption{Testing accuracy under IID distribution}
            \label{fig:test_acc_iid}
        \end{subfigure}

        \begin{subfigure}{0.9\linewidth}
         
            \includegraphics[width=\textwidth]{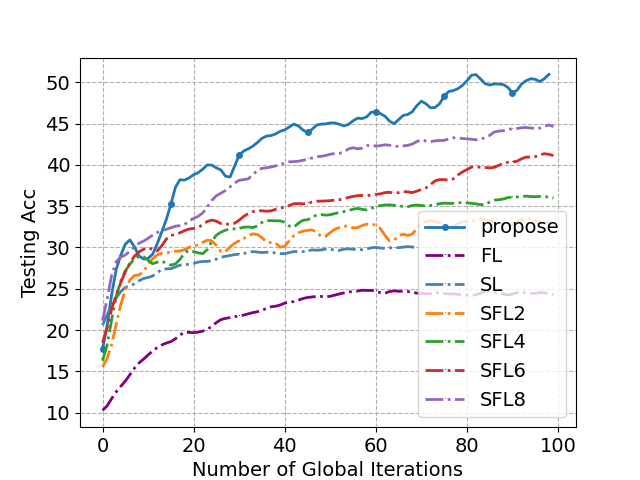}
            \caption{Testing accuracy under non-IID distribution}
               \label{fig:test_acc_non-IID}
        \end{subfigure}
    \end{minipage}
    \caption{Performance of case study.}
    \label{performance}
\end{figure*}

\subsubsection{Performance Analysis}
Fig. \ref{performance}(\subref{communication_overload}) shows that the communication load of different schemes with one local epoch and one round is decrease with the increase of the number of cut layer. As We can see, the load of SL and SFL is much more than that of FL. This is because the intermediate values calculated by the model need to be transmitted by the network.\par

Fig. \ref{performance}(\subref{communication_time}) shows the overall training time (including communication time and computation time) of different system design. The serial SL calculates and communicates with four vehicles in sequence, which consumes an additional significant amount of time. The proposed ASFL cost less time than FL and SL show that that the ASFL performance well. Although the communication loads of ASFL and SL are much higher than that of FL, it can be concluded from the experiment that the overall training time of ASFL is slightly less than that of FL, indicating that ASFV is using increased communication load to reduce computational load and finally reduce global training time is reasonable, which reflects the rationality of the ASFV architecture. \par

    Fig. \ref{performance}(\subref{fig:test_acc_iid}) shows the testing performance under IID data distribution. As we can see, the SFL schemes have better performance than SL and FL. Surprisingly, we found a correlation between the model performance of the SFL scheme and the choice of cut layers, with the performance improving as we chose the later cut layers. Fig. \ref{performance}(\subref{fig:test_acc_non-IID}) shows the testing performance under non-IID distribution, which every vehicle only choose six classes data out of ten. The SL has better performance compared with FL. Notably, our proposed ASFL scheme outperforms other alternatives. \par

\section{Open Research Directions}
    SFL, serving as distributed learning framework, have attracted significant attention, yet the research remains in its early stages. Especially within the domain of vehicular networks, there are quite a few research directions awaiting for further investigation.

\subsection{Data Generation and Selection}
    The advantage of AI lies in using a large amount of local device data for training. However, in SFL, while the system can access massive data from numerous devices, these data are commonly non-IID, resulting in limitations in the learning and generalization ability of the system model. With the development of Artificial Intelligence Generated Content (AIGC) technology, we consider using it to assist in generating data to mitigate the impact of non-IID data distribution on the model, thereby improving the performance of the system. The use of data generation for training raises three key issues. Firstly, how to evaluate or measure the effectiveness of data. {Secondly, large-scale training datasets often contain invalid or redundant data. Selecting relevant data is crucial to reducing communication and computational burden.} Thirdly, how to balance data generation and model training performance in mobile scenarios of vehicle networks.
\subsection{Cut Layer Selection }
    In SFL, the global model is divided into non-overlapping vehicle-side models and RSU-side models through cut layer. The latter the cut layer is chosen, the smaller the size of the smashed data. As vehicle speed increases, the channel stability between vehicles and RSU weakens, thus selecting cut layers latter to reduce communication load. However, with the acceleration of vehicle speed, the time spent by vehicles within the RSU communication range decreases. Therefore, selecting cut layers former reduces vehicle-side model computation time. Thus, designing a cut selection strategy to balance vehicle communication and computation resources, as well as time-energy load, to achieve minimal overall training time is a significant consideration.\par
    
    Cut layer selection strategies need to consider not only the balance between computation and communication but also the balance between privacy protection and cost. The latter the cut layer, the greater the computational load on vehicles, the smaller the communication load, and the better the privacy of the smashed data (the output smashed data will be more blurred). Therefore, it is necessary to consider how to balance communication load and privacy for vehicles with different velocity and capabilities.

\subsection{Split Inference}
As Transformer, AIGC, and LLM technologies advance, an increasing number of intelligent vehicular networking services rely on their support. However, directly deploying large-scale models for training on vehicles is impractical. Firstly, vehicles lack sufficient computing resources to support such tasks. Secondly, running large models on vehicles consumes excessive energy, thereby impacting vehicle transportation time. Therefore, employing the concept of distributed learning, known as split learning, to decompose Transformer architecture models into what is termed split inference, has become a noteworthy research direction. This approach can be applied in vehicular networking environments assisted by AIGC and LLM to reduce demands on vehicle-side resources while maintaining system performance and efficiency.\par
Distinguishing split inference from split learning is crucial. In SL, the outputs of the cut layer (smashed data) are shared during forward propagation, while only the gradients from the smashed data are transmitted back to the vehicle during back propagation. In contrast, split inference involves sending the outputs of the cut layer to the server without requiring back propagation.\par

\subsection{Wireless Resource Allocation}   

When discussing distributed learning frameworks, it is essential to use wireless networks to transmit a large amount of intermediate training data. However, this approach also creates significant communication overhead. Considering a large number of vehicles involved in each training iteration, it is essential to allocate communication resources wisely and effectively to speed up the implementation of VEI. Multi-objective optimization can be comprehensively explored by combining variables such as delay, energy consumption, convergence time, and learning accuracy, thereby promoting the rational resource allocation. In addition, designing incentive mechanisms for resource allocation is also a means of promoting rational resource allocation. When formulating incentive plans, not only must factors such as CPU/GPU frequency, spectrum, and energy costs be considered, but the impact of transmission interference on other vehicles can also be considered as a cost factor. Various schemes based on game theory, contract theory, and auction theory are also useful in ASFL incentive design within vehicular networks.

\subsection{Parallel Design}
    In vehicular networks, the number of vehicles within the server's communication range constantly changes due to their mobility. When designing optimization algorithms and incentive mechanisms, it's important to consider not only minimizing system load or maximizing model performance but also ensuring system scalability. When many vehicles involve in the network, a reasonable and high-efficient parallel design can help to avoid a linear or exponential increase in system latency, ensuring strong scalability of the system.

\section{Conclusion}
{This paper provides a comprehensive review and introduction to the concept of Vehicular Edge Intelligence (VEI), including centralized and distributed architectures, performance metrics, and associated challenges. In recent years, distributed machine learning methods, such as Federated Learning (FL), Split Learning (SL), and Split Federated Learning (SFL), have been widely applied in vehicular networks to support VEI. However, these methods face significant challenges due to the inherent mobility and resource constraints of vehicular networks. To address these challenges, we propose an enhanced Adaptive Split Federated Learning (ASFL) scheme designed to overcome the limitations of traditional SFL in VEI environments. This scheme optimizes system performance through a mobility-adaptive cut layer selection strategy. Our case study highlights the significant advantages of the ASFL approach, particularly in improving model performance with non-IID data, where it demonstrates superior test accuracy and training latency compared to FL, SL and SFL, thus showcasing the potential of ASFL. Finally, we conduct a thorough analysis of the ASFL design and  suggest potential directions for future research.}


\bibliographystyle{IEEEtran}
\bibliography{ref}

\end{document}